\documentclass[%
 aip,
 amsmath,amssymb,
 reprint,%
]{revtex4-1}

\newcommand{\suppl}{Supporting Information }

\usepackage{graphicx}
\usepackage{dcolumn}
\usepackage{bm}

\usepackage[utf8]{inputenc}
\usepackage[T1]{fontenc}
\usepackage{mathptmx}
\usepackage{etoolbox}

\usepackage{xcolor}
\usepackage{ulem}
\usepackage{bm}

\makeatletter
\def\@email#1#2{%
 \endgroup
 \patchcmd{\titleblock@produce}
  {\frontmatter@RRAPformat}
  {\frontmatter@RRAPformat{\produce@RRAP{*#1\href{mailto:#2}{#2}}}\frontmatter@RRAPformat}
  {}{}
}%
\makeatother
\begin{document}

\preprint{AIP/123-QED}

\title[Transverse Inscription of Silicon Waveguides by Picosecond Laser Pulses]{Transverse Inscription of Silicon Waveguides by Picosecond Laser Pulses}
\author{Markus Blothe*}
\email[]{markus.blothe@uni-jena.de}

\author{Alessandro Alberucci}
\author{Namig Alasgarzade}
\author{Maxime Chambonneau}

\affiliation{ 
Friedrich Schiller University Jena, Institute of Applied Physics, Abbe Center of Photonics, Albert-Einstein-Straße 15, Jena 07745, Germany
}%

\author{Stefan Nolte}
\affiliation{ 
Friedrich Schiller University Jena, Institute of Applied Physics, Abbe Center of Photonics, Albert-Einstein-Straße 15, Jena 07745, Germany
}
\affiliation{%
Fraunhofer Institute for Applied Optics and Precision Engineering IOF, Center of Excellence in Photonics, Albert-Einstein-Straße 7, Jena 07745, Germany
}%

\date{\today}
 
\begin{abstract}

In this paper, picosecond laser inscription of segmented waveguides in crystalline silicon based on a deterministic single-pulse modification process is demonstrated. Pulses of 43 ps duration at 1.55 µm wavelength are used to transversely inscribe periodic structures with a pulse-to-pulse pitch of around 2~µm. Infrared shadowgraphy images and Raman spectroscopy measurements indicate that the modifications exhibit a spherical shape. Characterization of waveguide performance at 1.55~µm for various pulse energies and periods is carried out. Direct comparison with numerical simulations confirms the presence of graded index waveguides, encompassing a micrometer core size and a maximum refractive index change of around $7\times 10^{-3}$. This short-pulse inscription approach can pave the way for three-dimensional integrated photonic devices in the bulk of silicon.
\end{abstract}

\maketitle
 \section*{Keywords} Silicon, waveguide, photonics, ultrashort pulses, laser material modification.

\section{\label{sec:Intro}Introduction}
Silicon is by far today's most important material for micro-electronics. Silicon-based electronic devices grew exponentially in their data processing capabilities and compactness in the last sixty years, although doubts about the validity of Moore's law in the near future recently emerged.\cite{Moore1965}  
To maintain the pace dictated by this empirical law, integration of photonics with electronics has been proposed to achieve faster and more efficient data transfer inside multi-core processors, or between processors on the same board. The field of silicon photonics is predicted to bring major advantages in this regards by providing monolithic optical components integrated directly into the electronic chips.\cite{Lipson2005,Baehr-Jones2012,Dong2014,Clark2021} 
Data transfer can then be performed optically via waveguides integrated into the chips themselves, without resorting to external elements such as optical fibers. For the sake of compatibility with current silicon factories, silicon waveguides are so far mostly based on Silicon-on-Insulator (SOI) technology, that is, an essentially planar lithography process which is strongly limited in terms of full three-dimensional (3D) integration.

One possibility to inscribe waveguides in a fully 3D way, as needed also for the next steps of optical neural networks,\cite{Shen2017} is by direct laser writing (DLW), which is extensively studied for dielectrics and other wide-bandgap materials. This technique relies on high intensity laser pulses with a wavelength in the transparency range of the material which are tightly focused into the bulk. Nonlinear absorption takes place locally in the high intensity regions, without damaging the entrance surface. This allows to induce localized refractive index modifications in a precise way, leading to the realization of a plethora of optical components within the last three decades, ranging from waveguides to diffraction gratings and wave-plates based upon nanogratings.\cite{Davis1996,Gattas2008,Ams2017} Also segmented waveguides were realized which consist of a periodic structure of refractive index modulation and behave similar to a continuous refractive index modulation with an average $\Delta n$. They are first discussed in Ref.10~\cite{Weissman1993} and then applied to the direct laser writing case to combine waveguiding with Bragg reflection.\cite{Ams2017}

First results to transpose this DLW technique to silicon were reported nearly two decades ago where waveguides are transversely written close to an Si-SiO$_2$ interface with femtosecond pulses.\cite{Nejadmalayeri2005} However, it was proven in recent years that the actual modification of the bulk of silicon with femtosecond laser pulses is a challenging task due to nonlinear propagation effects.\cite{Chanal2017, Chambonneau2021a} The combination of Kerr effect, multiphoton ionization and plasma defocusing leads to filamentation and with this to intensity clamping, limiting the amount of deposited energy in the volume to levels below the modification threshold.\cite{Chambonneau2021b, Chambonneau2023}  
A simple method to reduce these nonlinear influences and enable deterministic bulk modification is the use of longer pulses.\cite{Das2020} In the nanosecond case, more recent progress has been reported for the inscription of gratings \cite{Chambonneau2018,sabet2023}, waveplates\cite{Saltik2024}, and waveguides. The waveguides were written in two configurations: along the optical axis (longitudinal inscription)\cite{Chambonneau2016,Tokel2017,Turnali2019} and perpendicular to it (transverse inscription).\cite{Wang2020,Wang2021}
While the use of nanosecond laser pulses eases up the difficulties connected to nonlinear propagation in silicon because of the reduced peak intensities, it is expected that due to the longer laser-material interaction a larger heat-affected zone (HAZ) forms, accompanied with shock waves, which lead to the formation of voids and cracks.\cite{Gattas2008,Verburg2014a}

Due to the limitations associated with high-intensity pulses, femtosecond inscription of waveguides in silicon has been demonstrated only longitudinally by surface seeding so far. For this, the decreased modification threshold of the backside interface enables the modification of a small volume.\cite{Matthaus2018,Alberucci2020} The locally altered material features a lower modification threshold, thus acting like a seed for the subsequent laser pulses, eventually allowing the inscription of continuous structures in a longitudinal way.\cite{Pavlov2017,Matthaus2018,Alberucci2020} On the other hand, this seeding has limited the transverse inscription of overlapping structures so far.\cite{Chambonneau2021} However, longitudinal inscription comes with different drawbacks. The produced structures are of limited length, dictated by the sample thickness and by the working distance of the focusing optics. Further, due to the high refractive index of silicon, the intensity distribution in the focal region is strongly dependent on the focusing depth if no sophisticated countermeasures such as aberration correction or depth-dependent energy adjustment are taken. 

Taken together, transverse waveguide inscription with sub-nanosecond laser pulses would represent a critical step towards laser-inscribed silicon photonic devices.
Picosecond pulses might proof as a beneficial middle-ground for the pulse duration as they provide the possibility of deterministic volume modification and can still show a reduced HAZ in comparison to nanosecond pulses.\cite{Das2020} 

In this paper, we present transverse inscription of segmented waveguides in silicon based on a single-pulse modification process with picosecond laser pulses.  
First, the modification features are analysed by infrared transmission microscopy and Raman spectroscopy. The waveguide performance versus the writing energy is explored afterwards by end-fire coupling 1.55~µm radiation into the structure and measuring the output field. 
Refractive index maps and effective refractive indices for the waveguides are then deduced from the near-field images. The experimental results are compared with numerical simulations to confirm the nature of the waveguiding and the associated refractive index landscape. 
Finally, the monomodal nature of the waveguides is experimentally demonstrated by lateral shift experiments. \cite{Alberucci2020}

\section{\label{sec:Characterization_Crosssection}Characterization of modifications}
\begin{figure}
\includegraphics[width=8.5cm]{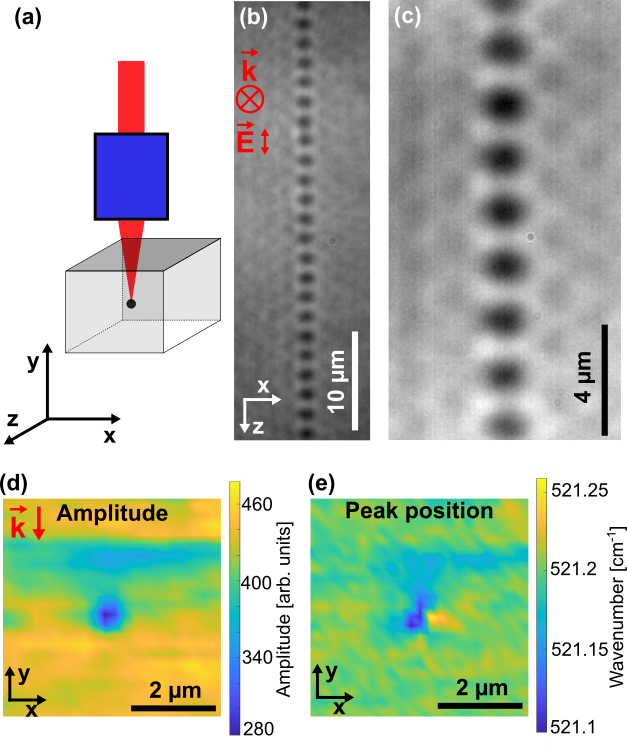}
\caption{\label{fig:tm_image} (a) Schematic of the inscription process with the coordinate system used throughout the paper. (b-c) Shadowgraphic top-view images of typical modifications written with a pulse energy $E_{\mathrm{in}} = 1.25$~µJ at two different magnifications identified by the scale bar in the image. The vectors $\bm\vec{k}$ and $\bm\vec{E}$ indicate the inscription beam propagation direction and its polarization direction, respectively. (d) Amplitude and (e) peak position of the Raman signal at 521~cm$^{-1}$ mapped across the cross-section of the waveguide.}
\end{figure}

For the inscription of modifications, a laser source with wavelength $\lambda$~=~1.55~µm and pulse duration $\tau$~=~43~ps was used. The beam was focused 200~µm below the surface of a crystalline silicon sample by an aspheric lens with a numerical aperture NA~=~0.85. For more details on the inscription see the Methods section as well as the Supporting Information, Section I. The inscribed sub-surface structures have been first studied by means of standard microscopy methods using a wavelength $\lambda$ = 1.317~µm and a microscope objective with NA~=~0.85. The Abbe limit thus provides  $d=1.22 \lambda / \text{NA} = 1.9$~µm for the optical resolution.

Shadowgraphy images of the $xz$-plane (see Figure~\ref{fig:tm_image}~(a) for the orientation) provide a clear and detectable signal and a typical portion of the observed structure written for $E_{\mathrm{in}} = 1.25$~µJ pulse energy is shown in Figure~\ref{fig:tm_image}~(b) and (c) with different scales. Periodic black circular zones arranged in a "pearl necklace"-like structure can be observed. From Figure~\ref{fig:tm_image}~(c) one can deduce that the individual black zones do not overlap with each other. Their diameter is thus slightly below 2~µm, the latter corresponding to the pulse-to-pulse pitch determined by our scanning velocity and laser repetition rate. Hence, our structures closely recall segmented waveguides already realized in glass and wide-bandgap crystals using femtosecond writing.\cite{Ams2017} Due to the size of the modifications, a relevant diffraction pattern is expected, explaining the periodic intensity modulation besides the structure. Such strong diffraction also inhibits the direct full 3D reconstruction of the optical properties along the $y$ direction. An almost circular shape of the single-pulse modifications is observed for linear polarization along the inscription direction, in contrast to the multi-pulse irradiation regime.\cite{Mori2015,Das2020,A.Wang20201,Chambonneau2021} With increasing pulse energy, the absorbing zones undergo a small growth and the contrast slightly improves as well. See \suppl Figure~S3, Section II for this and modifications inscribed with linear polarization perpendicular to the inscription direction as well as circular polarization.

Modification tracks were not unambiguously detectable by shadowgraphy on the $xy$-plane, corresponding to the cross-section of the waveguide, because of partly guided light as well as the influence of the other modifications along the probing direction. Further, visible light microscope imaging in reflection does not provide a contrast at the cross-section surface.
Therefore, the modifications were further characterized using Raman spectroscopy on the $xy$-plane.

As shown before, laser-induced modifications lead to a clear amplitude drop for the main silicon peak located at 521~cm$^{-1}$ and a shift of the position of this peak, ascribable to a perturbed crystal lattice and optically-induced strain in the material, respectively.\cite{Mori2015,Kammer2019,Blothe2022} Both effects can be observed in the amplitude and peak position maps shown in Figure~\ref{fig:tm_image}~(d) and (e). Interestingly, the modification appears circular in the amplitude map, so as in the $xz$-plane. A precise determination of the size in the cross section is hindered by the modulation of the modification along the $z$ direction, plus the fact that Raman measurements at 532~nm only probe a limited depth of a few nanometers in silicon. Nonetheless, a diameter on the plane $xy$ comprised between 1~µm and 2~µm can be estimated. The shifted position of the peak shown in Figure~\ref{fig:tm_image}~(e) suggests that the strain induced in the material might actually be responsible for the induced refractive index change shown below. \cite{Munguia2008,Cai2013}

\section{\label{sec:Characterization_WG}Determining the waveguide properties}

In the paraxial regime and neglecting back-reflections, the slow envelope $A$ of a monochromatic wave at wavelength $\lambda$ propagates according to the paraxial Helmholtz equation
\begin{equation}  \label{eq:paraxial}
    2ik_0 n_0 \frac{\partial A}{\partial z} + \nabla^2_t A + k_0^2 \left( n^2(x,y,z) - n_0^2 \right) A =0,
\end{equation}
where $\nabla^2_t=\partial^2_x+\partial^2_y$ is the transversal Laplacian, $n_0\approx 3.5$ is the unperturbed refractive index of silicon, and $k_0=2\pi/\lambda$ is the vacuum wave number. Hereafter, we will fix $\lambda=1.55~$µm, the latter being the wavelength employed in all characterization experiments. The waveguide profile is provided by the light-induced permanent changes in the refractive index profile, $n^2(\bm r)-n_0^2\approx 2n_0 \Delta n(\bm r)$ due to the small index contrast associated with laser-written waveguides. As shown in Figure~\ref{fig:tm_image}~(b) and (c), the induced perturbation in the refractive index is periodic with period $\Lambda$. For the sake of simplicity, we consider the periodic part to be sinusoidal along $z$, if not otherwise stated. We also assume the transverse profile to be the same for the continuous and periodic terms, $\Delta n(\bm r)= V(x,y)\left[1 + \sigma \sin\left(\frac{2\pi z}{\Lambda}\right)\right]$, where $\sigma$ determines the relative weight of the two contributions.  For our purpose, the guided modes in segmented waveguides can be found by simply considering the averaged value of the refractive index along $z$ (see Supporting Information, Section VI for a detailed discussion). Although the fast varying scales can affect the averaged intensity through the so-called Kapitza effect, such effects are negligible for the range of parameters of our waveguides.\cite{Alberucci2013} The $lm$-th quasi-modes $A_{lm}(\bm r)= \psi_{lm}(x,y) e^{i\beta_{lm} z} e^{i \sigma \left(\Lambda/\lambda\right) V(x,y)\cos\left( 2\pi z/\Lambda\right)}$ then satisfy the eigenvalue problem
\begin{equation} \label{eq:modes}
     \beta_{lm} \psi_{lm} = \frac{1}{2 k_0 n_0} \nabla^2_t  \psi_{lm} + \frac{k_0}{2} V(x,y)  \psi_{lm},
\end{equation}
where $\beta_{lm} = n^{(lm)}_\mathrm{eff}k_0$ is the propagation constant, being $n^{(lm)}_\mathrm{eff}$ the associated effective index. The fundamental mode is labelled with $l=0,m=0$. 
Waveguides can then be characterized by finding the profile $\psi_{lm}$ of all the guided modes plus their effective refractive indices. Hereafter we will consider only monomodal waveguides, thus we will drop the indices $l$ and $m$ in the remainder of the paper.

Experimentally speaking, the waveguides are characterized by recording the near-field images (thus measuring $|\psi|^2$), then extracting the effective index through the exponential tails. As it will be shown below, the observed guided modes feature cylindrical symmetry and pronounced exponential tails: on one hand, symmetry will be accounted for defining the transverse radial distance $r=\sqrt{x^2+y^2}$ with respect to the center of the waveguides; on the other hand, tails make a standard Gaussian fit unsuited in our case. Hence, the intensity cross-sections will be fitted by the sum of a generic Gaussian function 
\begin{equation} \label{eq:simple_gaus}
I_G = I_0\exp\left[-2(r-\bar{r})^2/w^2\right],
\end{equation}

and its convolution with a double exponential decaying function 
\begin{equation} \label{eq:double_exp_decay}
R=R_0\exp(-\lvert r \rvert/l).
\end{equation} The rationale is that the Gaussian function and the bilateral exponential account for the central portion of the mode (that is, the portion overlapping with the core of the waveguide) and the exponential tail region where $\Delta n\approx 0$, respectively (see also the Methods section).

 Another relevant parameter is the amount of losses sustained by light when propagating through the waveguides, given that the intense writing laser unavoidably generates additional defects in the material. Experimentally, a simple method consists of insertion loss measurement by determining the input and output powers. To specify the propagation loss, the waveguide length and different contributions to the overall losses needs consideration. With respect to the waveguide interfaces, Fresnel losses occurring at the two air-Si interfaces, amounting to a loss of 3.18~dB under the assumption of normal incidence. At the input interface, in the case of waveguiding, the amount of light coupled to a guided mode depends on the similarity between the input beam (let us call it $\varphi(x,y)$) and the guided mode $\psi$, as determined by the scalar overlap integral \[c = \int{\varphi \psi^* dxdy}\left/\left(\sqrt{\int{|\varphi|^2 dxdy}}\sqrt{\int{|\psi|^2 dxdy}} \right)\right ..\] With respect to the bulk of the waveguides, linear losses are neglected because silicon is highly transparent at $\lambda=1.55~$µm.\cite{Degallaix2013} With the former assumptions, the propagation losses owed to the defects induced by the laser-writing procedure can then be evaluated from the experimental data.

\subsubsection{\label{sec:Energy_dependence}Energy dependence}
\begin{figure*}
\includegraphics[width=17.8cm]{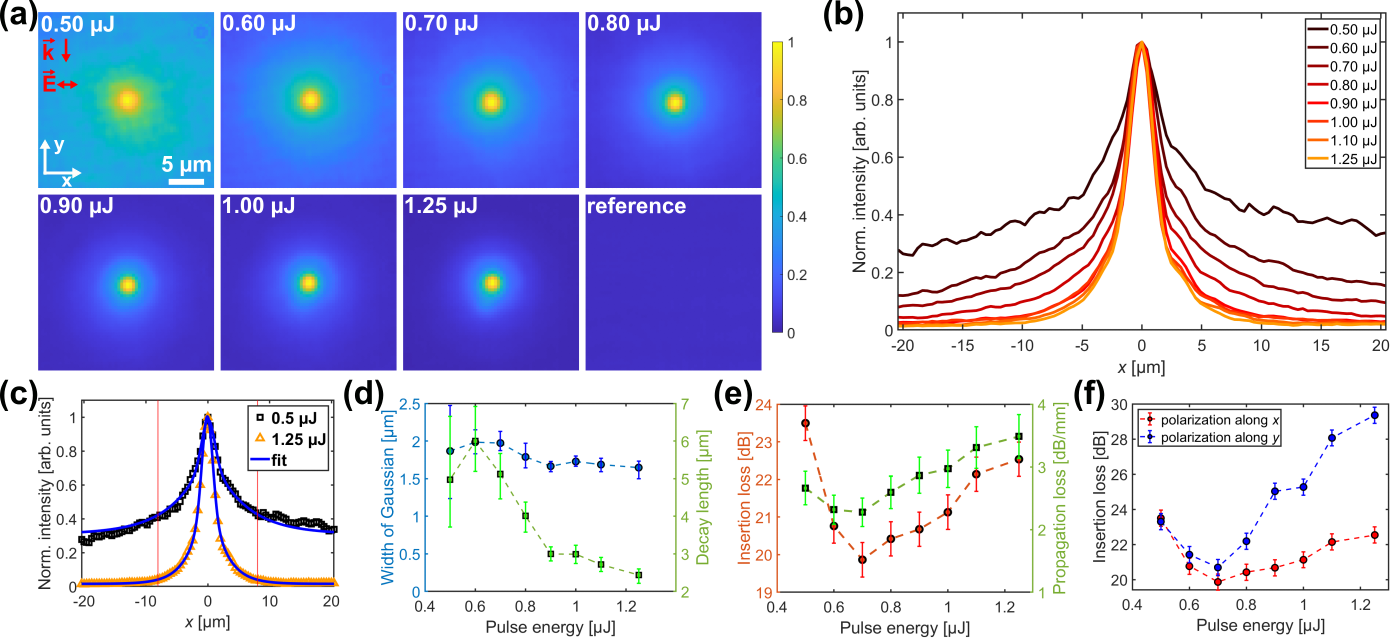}
\caption{\label{fig:wg_performance} Recorded near-field images and derived data for waveguides inscribed with different inscription energies. (a) Normalized near-field images for the indicated inscription pulse energy and a pristine reference position, the latter normalized to the case $E_\mathrm{{in}}=1.25~$µJ. The laser propagation direction for the inscription and the linear probing light polarization are indicated with $\vec{k}$ and $\vec{E}$ respectively. (b) Normalized intensity cross sections in $x$ direction. (c) Two intensity cross-sections at the indicated inscription energies with the corresponding fit functions (blue). (d) Radius of the calculated averaged ($x$ and $y$ direction) Gaussian width (blue) and exponential decay constant (green) for all waveguides derived from the fit. The error bars represent the 95\% confidence interval of the fits in both directions. (e) Insertion (red) and propagation loss (green) in dependence of inscription energy for examination polarization along $x$. The error bars represent a $\pm$ 10\% deviation in the measured intensity. (f) Insertion loss for two perpendicular examination polarizations.}
\end{figure*}

Figure~\ref{fig:wg_performance}~(a) shows normalized near-field images for 4.5~mm long waveguides written 200~µm below the surface for different pulse energies. The scanning velocity and repetition rate are adjusted to achieve a 2~µm pulse-to-pulse pitch. The probe beam employed to characterize the waveguides is Gaussian-shaped ($\varphi=E_0e^{-r^2/w_\mathrm{in}^2}$) with a waist radius $w_\mathrm{in}$ $\approx$ 2.4~µm, corresponding to a Rayleigh distance of around 40~µm. The polarization of the probe beam is horizontal (i.e., parallel to $x$-axis). Vertical polarizations show very similar profiles (see the Supporting Information Figure S4, Section III). As mentioned above, the output profiles are radially-symmetric within a good accuracy, showing average deviations of $\approx$10\%, and $\approx$15\% for the width and decay length, respectively.
For all cases, the output beam is much narrower than in the homogeneous material case, shown in the panel labeled reference, which is normalized with respect to the $E_\mathrm{{in}}=1.25~$µJ case. The beam width in the free diffracting case is approximately 270~µm, thus demonstrating at least the presence of a focusing structure as an intense central lobe is indeed always observed. For low inscription energies ($E_\mathrm{{in}}<0.8~$µJ), the central peak is surrounded by a broad radiation background, exponentially decaying with the distance from the center of the waveguide $r$. The increasing confinement with inscription energy is clearly illustrated in Figure~\ref{fig:wg_performance}~(b), where the evolution of the normalized intensity cross-section for different input energies is shown. Furthermore, the waveguides are verified to be monomodal by shifting the input beam along the transverse plane $xy$ (see Section~\ref{sec:shift} for further details).
To quantify the confinement as a function of the writing energy $E_\mathrm{{in}}$, we fit the acquired intensity profile with the before mentioned functions given in Equations~\eqref{eq:simple_gaus} and \eqref{eq:double_exp_decay}. The similarity between the fitting function and the measured profile is always very good. Figure~\ref{fig:wg_performance}~(c) shows the comparison for the cross-section of the lowest ($E_\mathrm{{in}}=0.5~$µJ) and the highest ($E_\mathrm{{in}}=1.25~$µJ) energy with their corresponding fits (blue line). The retrieved width of the central Gaussian part $w$ (blue) and the exponential decay length $l$ (green) are plotted as a function of $E_\mathrm{{in}}$ in Figure~\ref{fig:wg_performance}~(d). In agreement with the qualitative statements made earlier, both fitting parameters $w$ and $l$ reduce with increasing writing energy, thus confirming an improvement in the confinement as the pulse energy ramps up. The upper limit in $E_\mathrm{{in}}$ is fixed by the formation of double-humped waveguides, possibly caused by the formation of two distinct focal regions due to larger nonlinear effects occurring during the inscription process. 
Figure~\ref{fig:wg_performance}~(e) shows on the left axis how the insertion loss (red circles) depend on the inscription energy. It reaches a minimum of $\approx$ 20~dB for a pulse energy of 0.7~µJ. The corresponding propagation loss (green squares) on the right axis, with consideration of the Fresnel losses, the mode overlap, and the length of the waveguide, range from 2.3 to 3.5 dB/mm and show a non-monotonic behavior with a similar trend as the insertion loss. 
The presence of a local minimum will be explained below by numerical simulations. On the other side, the continuous increase in losses for $E_\mathrm{{in}} > 0.7$~µJ is mainly associated with an increase in the density of scattering centers due to the higher intensities. When the probe is vertically polarized, the losses at low energies are comparable with respect to the horizontal input, as can be seen in Figure~\ref{fig:wg_performance}~(f). The transmission is instead much worse for high $E_\mathrm{{in}}$, resulting at $E_\mathrm{{in}}=1.25~$µJ in an additional maximum loss of 6.8~dB, i.e., around 80\% less power than for horizontal polarization. This large polarization dependence can be partly ascribed to the strong anisotropy of the scattering centers, a property confirmed by the strong dependence on the observation direction for the light scattered in the directions normal to the waveguide axis $z$.\cite{Monier:2010} Further, birefringence deduced from slightly higher calculated refractive indices might also contribute to the difference in transmission. The similar losses at low energy suggest that scattering centers play a minor role in this regime and the overall best waveguides in terms of losses can be obtained here, as long as the refractive index profile is able to support a mode of the desired width.

A fundamental quality parameter is the repeatability of the inscription process, a critical issue for longitudinal inscription schemes \cite{Alberucci2020} and, in a broader view, a potential critical point due to the nonlinear optical effects employed in the fabrication. For this purpose, we tested five waveguides inscribed in the same sample with identical parameters, which are 0.9~µJ pulse energy and 2~µm pulse-to-pulse pitch. Very similar results are observed for the intensity profiles and transmitted powers measured for different realizations, with maximal variations of 3\%, while our characterization setup has an estimated accuracy of around 1\%.
Slightly higher variations were observed for waveguides inscribed in different samples with the same parameters, highlighting the importance of rigorous alignment procedures during inscription and characterization.
Summarizing, our transverse-writing procedure features a good repeatability, in contrast to the longitudinal scheme.\cite{Alberucci2020} This net improvement can be associated with the single-pulse modification in each point, thus avoiding the interaction of an intense pulse with the temporary and permanent modifications introduced by previous pulses.

In the next step we aim to extract information on the refractive index shape $V(x,y)$ of the laser-inscribed waveguides from the observed
output profiles. The acquired output intensity profiles are all featuring well pronounced exponential tails, thus permitting an accurate measurement of the decay length. For $V\approx 0$, Equation~\eqref{eq:modes} under the radial symmetry assumption turns into a modified Bessel equation.\cite{Yariv1997} Thus, the mode far enough from the core can be described by $K_0 (\alpha r ) \sim r^{-\frac{1}{2}}e^{-(\alpha r)}$; $K_0$ is the 0-th order modified Bessel function of the second kind, and we also used the asymptotic form of $K_0$ for large $r$. The effective refractive index $n_\mathrm{eff}$ is thus related to the exponential decay constant $\alpha$ by  \cite{Yariv1997,Alberucci2020}
\begin{equation} \label{eq:alpha_neff}
\frac{\alpha}{k_0} \approx \sqrt{n_\mathrm{eff}^2 - n_0^2}.    
\end{equation} 

\begin{figure}
\includegraphics[width=8.5cm]{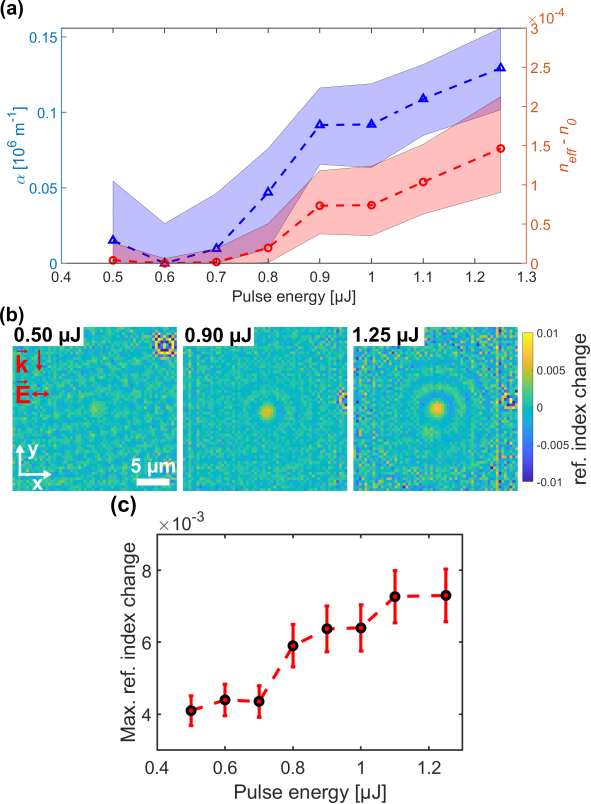}
\caption{\label{fig:dn} (a) Exponential decay constant $\alpha$ (blue triangles) and corresponding calculated effective refractive index $n_\mathrm{eff}$ (red circles) for different inscription energies. The mean value of a fit starting at $r=$5~µm in $x$ and $y$ direction are connected by the dashed lines. The shaded region around these marks the 95\% confidence interval for fits starting in the interval 3.5$<r<$8~µm. (b) Reconstructed refractive index maps by the inversion of the Helmholtz equation for the indicated inscription energies. The vectors $\vec{k}$ and $\vec{E}$ indicate the inscription beam propagation direction and linear examination polarization, respectively. (c) Retrieved maximum refractive index change in dependence of the inscription energy. The error bars represent a $\pm$10\% error, estimated from the observed changes for different realizations with same parameters.}
\end{figure}

An estimate for $\alpha$ is retrieved indirectly from the acquired data by fitting the function $e^{-2\alpha r / r}$ for large enough $r$ to the fit of the intensity profile described earlier in $x$ and $y$ directions. A direct fit to the experimental data is less accurate due to the high noise present in the images, particularly degrading the signal-to-noise ratio (SNR) in the low-intensity regions. Empirically, we find that the best choice for a starting point in terms of radius $r$ is in our case in the interval 3.5~µm$~<r<$~8~µm, the highest starting value indicated in Figure~\ref{fig:wg_performance}~(c) by the red vertical line. The fit is then carried out from the starting point to increasing $r$, limited by the examined region (20~µm). A dependency of the retrieved data of the starting point is observed and described in more detail in the \suppl Figure~S6, Section V. The values of $\alpha$ can be seen in Figure~\ref{fig:dn}~(a), where the shaded area corresponds to the 95\% confidence intervals of the fit for starting points ranging from 3.5~µm to 8~µm in $x$ and $y$ direction.
For $E_\mathrm{{in}}>0.8~$µJ, a monotonic growth of $\alpha$, and thus $n_\mathrm{eff}$ [see Equation~\eqref{eq:alpha_neff}], takes place with pulse energy. A maximum $n_\mathrm{eff}$ of around $1.5 \times 10^{-4}$ is observed for 1.25~µJ pulse energy. The almost vanishing values of $\alpha$ for $E_\mathrm{{in}} < 0.8~$µJ fall together with a drastic decrease of the insertion loss shown in Figure~\ref{fig:wg_performance}~(e). Nonetheless, in this interval the fitting procedure is less accurate due to the non-negligible radiation modes.

To physically interpret the values of $n_\mathrm{eff}$, we need to retrieve the maximum refractive index change in order to understand how close to the cut-off condition the observed modes are. To see if any cut-off is there, see the theoretical treatment reported in the Supporting Information, Section VII. The scope can be achieved by inverting the Helmholtz equation~\eqref{eq:modes} once $|\psi|^2$ has been substituted with the experimental intensity distribution.\cite{Mansour1996} Such technique works assuming a flat phase front and monomodal waveguides.
Here we stress again that segmented waveguides behave like standard (i.e., invariant along $z$) waveguides, but featuring an index profile equal to the average along $z$. This means that in our experiments we have only access to the \textit{slow}, i.e. the average, component of the field (thus, $V(x,y)$), the longitudinal varying components being negligible due to the short modulation period along $z$. The reconstructed refractive index maps $V(x,y)$ are shown in Figure~\ref{fig:dn}~(b) for three writing energies. Given that the procedure requires a division by the intensity, the accuracy of the technique drops rapidly as the distance from the waveguide center $r$ increases. For energies where confinement is higher, the inversion procedure provides one or more rings of depressed index circumventing the central core. As discussed in the Supporting Information, Section X the multiple rings are an artifact coming from the finite bandwidth of the point-spread function associated to the imaging objective. Nonetheless, both a bell-shaped and a bell-shaped central peak surrounded by a depressed region waveguides provide a similar pattern of concentric rings. Due to our experimental accuracy, we cannot determine with certainty if the first ring is present in the material, but the experimental data suggest the presence of a depressed region surrounding the central lobe.

The maximum of the observed averaged refractive index change $V(x,y)$ as a function of the writing energy $E_\mathrm{{in}}$ is reported in Figure~\ref{fig:dn}~(c). The maximum of $V(x,y)$ is included in the interval between $3-4 \times 10^{-3}$ and $7-8 \times 10^{-3}$. Relative variations between waveguides written with the same parameters are below $10\%$. Figure~\ref{fig:dn}~(a) proves due to the smallness of $n_{eff} -n_0$ that the guided modes are close to the cut-off condition. Even more interestingly, the maximum of $V(x,y)$ as a function of the energy $E_\mathrm{{in}}$ and the effective index $n_\mathrm{eff}$ undergo an abrupt transition around $E_\mathrm{{in}}=0.7~$µJ, where a local minimum is observed for the insertion losses as well as propagation losses [see Figure~\ref{fig:wg_performance}~(e)]. This suggests a quite abrupt transition in the optical confinement around this inscription energy. There are two possible explanations for this behavior: i) the nonlinear inscription process induces a modification whose amplitude behaves non-monotonically with the input energy $E_\mathrm{{in}}$;
ii) at low energies the measured outputs do not correspond to the fundamental mode of the waveguides.
Hypothesis i) seems to be in contrast with the continuous and monotonic changes observed in the modes [see Figure~\ref{fig:wg_performance}~(b)]. 
If ii) is true, for $E_\mathrm{{in}}<0.7$~µJ the waveguiding effect is extremely weak, leading to the presence of long-lasting radiation modes overlapping with a very broad fundamental mode, thus invalidating both the inversion of the Helmholtz equation and Equation~\eqref{eq:alpha_neff}.
The procedure based on the overlap integral used to subtract the coupling loss from the insertion loss would not be correct either. This would also explain the non-monotonic trend in the propagation loss, seen in Figure~\ref{fig:wg_performance}~(e). 

To get a definitive answer to this question, we need to resort to numerical simulations for the waveguiding effect.
\begin{figure}
\includegraphics[width=.99\linewidth]{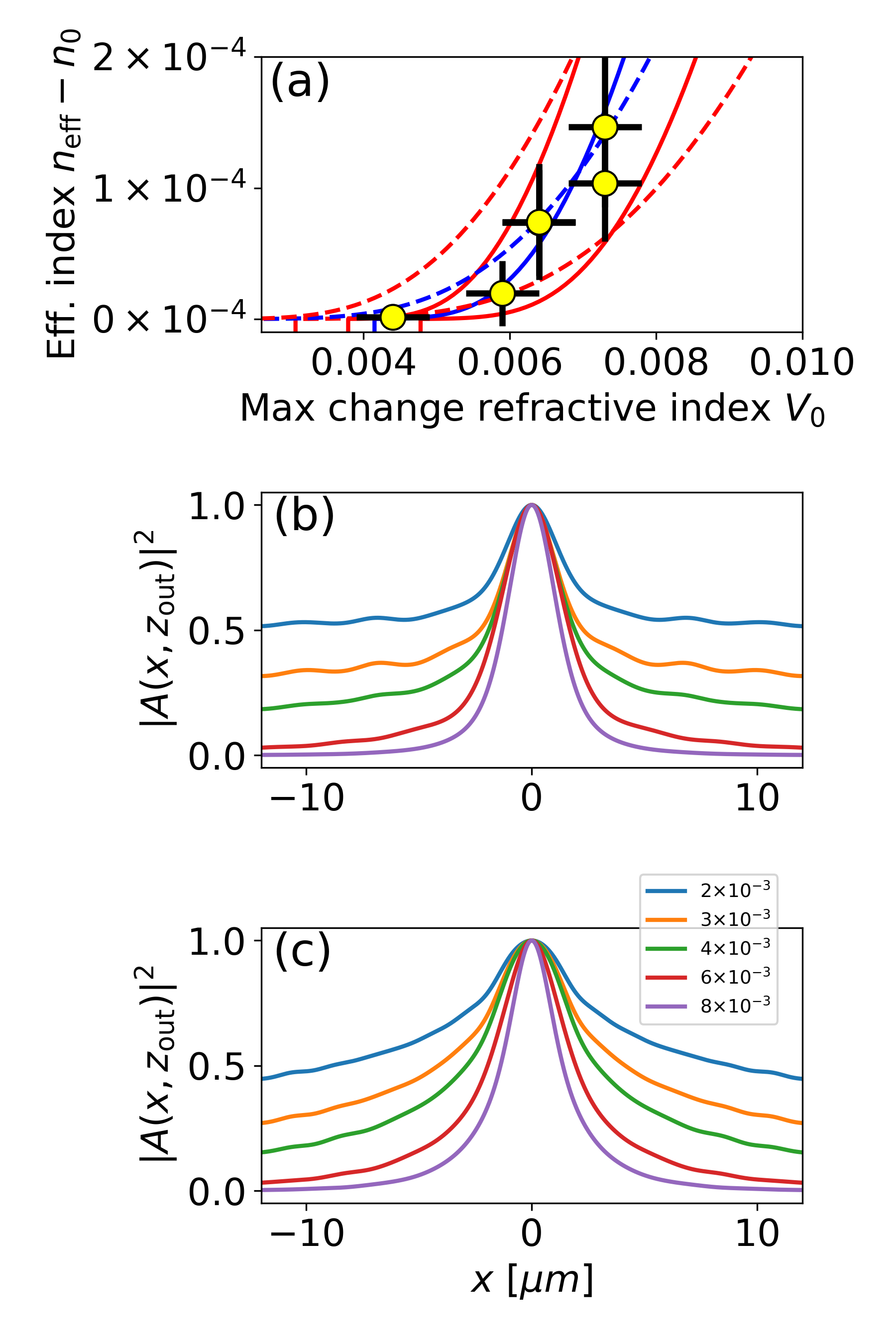}
\caption{\label{fig:theory_modes} (a) Effective index $n_\mathrm{eff}-n_0$ as a function of the amplitude of the waveguide $V_0$. Yellow symbols are the experimental data reported in Figure~\ref{fig:dn} satisfying the condition $E_\mathrm{{in}}\ge 0.7$~µJ (for the motivation of such a choice see the Supporting Information). Dashed and solid lines are the computed guided modes for Gaussian and W-shaped profiles, respectively. The blue curves correspond to $w_\mathrm{guide}=1.5~$µm and $w_\mathrm{guide}=2.85~$µm for Gaussian and W-shape respectively, whereas the red curves are the dispersion curves when variations of around $\pm~$150~nm are imposed to $w_\mathrm{guide}$ (refer to the \suppl for further details). (b-c) Intensity cross-sections sliced in $z_\mathrm{{out}}=4.5~$mm and normalized to their own peak computed by BPM simulations for (b) W-shaped with $w_\mathrm{guide}=2.85~$µm and (c) Gaussian waveguides featuring $w_\mathrm{guide}=1.50~$µm. Each color corresponds to a different $V_0$ whose values are labelled in the legend. }
\end{figure}
Given the cylindrical symmetry of the manufactured waveguides, we solved Equations~\eqref{eq:paraxial} and~\eqref{eq:modes} in log-polar coordinates (see Methods and the Supporting Information, Section VII for details). Following Ref.~27~\cite{Alberucci2020}, we made two different ansatz for the photonic potential $V(r)$
\begin{align}
    V_G(r) &= V_0 e^{-2r^2/w^2_\mathrm{guide}},  \label{eq:gaussian} \\
    V_W(r) &= V_0 \left(1-2\frac{r^2}{w^2_\mathrm{guide}} \right) e^{-2r^2/w^2_\mathrm{guide}}. \label{eq:Wshape}
\end{align}
The Gaussian shape depicted by Equation~\eqref{eq:gaussian} represents a standard bell-shaped distribution, whereas Equation~\eqref{eq:Wshape} is W-shaped, that means, a central lobe of higher refractive index surrounded by a ring of depressed index. Such W-shaped profiles have been observed for example in transversely-written waveguides in glass.\cite{Michele2019} In our theoretical treatment, we suppose that the shape of the waveguide does not change with pulse energy $E_\mathrm{{in}}$: only the amplitude of the waveguide $V_0=V_0\left({E_\mathrm{{in}}}\right)$ is assumed to depend on the intensity of the writing beam.

The effective index $n_\mathrm{eff}$ computed via the eigenvalue problem \eqref{eq:modes} and the corresponding experimental measurements are compared in Figure~\ref{fig:theory_modes}~(a). Both for simulations and experiments, the ratio $\left(n_\mathrm{eff}-n_0\right)/V_0$ is at its maximum around 0.02, which confirms that all the waveguides are very close to the cut-off condition due to the small ratio. As detailed in the Supporting Information, Section VII, the transverse size of the waveguides are compatible with the direct observation of the modification discussed in Figure~\ref{fig:tm_image}.
Figure~\ref{fig:theory_modes}~(b) and (c) show the intensity profiles $|A|^2$ computed with a beam propagation model (BPM) code from Equation~\eqref{eq:paraxial} at the end of the waveguide for W-shaped (b) and Gaussian (c) distributions. For both families, the intensity profile is in good qualitative agreement with the experimental results plotted in Figure~\ref{fig:wg_performance}~(b), including all the energies used for inscription. 

Summarizing, the numerical simulations confirm the presence of gradient index waveguides written by the picosecond laser, providing both effective indices and beam profiles compatible with the observations. The simulations also provide a clear explanation for the non-monotonic behavior of the insertion loss, the latter being related to the magnitude of the refractive index change and not to intrinsic properties of the modified material. Finally a better agreement is achieved when W-shape profiles are assumed (see \suppl for additional details), but the overall quality of the comparison of experiments with simulations does not permit to make definitive statements on the exact profile of the waveguides.

\subsubsection{\label{sec:period}Dependence on the period}

\begin{figure}
\includegraphics[width=6.5cm]{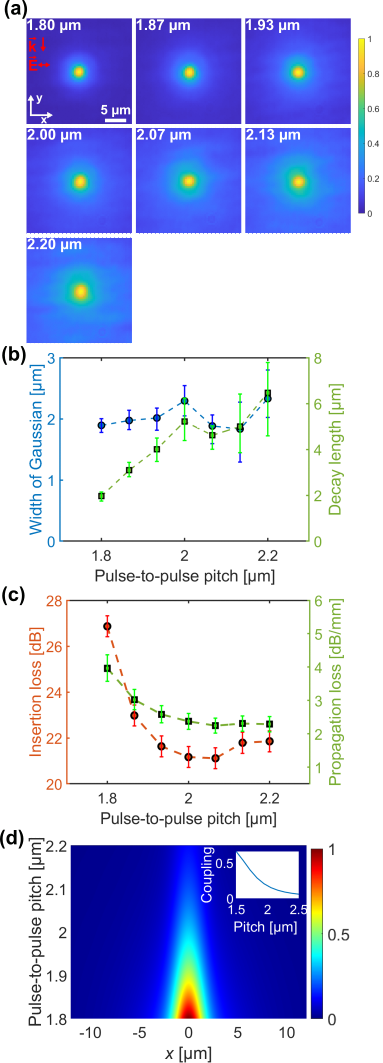}
\caption{\label{fig:wg_periodicity} (a) Normalized waveguide near-field images inscribed with the indicated periods for a constant pulse energy of 0.9~µJ. The inscription beam propagation direction is indicated with $\vec{k}$, the examination polarization direction with $\vec{E}$. (b)  Radius of the calculated Gaussian width (blue) and the decay length (green) of the fit function described in Section ~\ref{sec:Characterization_WG}. (c) Insertion loss and propagation loss.
(d) Cross-section of the output field calculated as a function of the pitch (vertical axis) via BPM simulations. The profiles are normalized with respect to the carried power. Inset: the corresponding coupling ratio as a function of the modulation period.}
\end{figure}

Let us now explore if the transverse confinement can be tuned by changing the distance between two consecutive modifications (i.e., the period $\Lambda$) instead of the inscription energy $E_\mathrm{{in}}$. 
Supposing two periods $\Lambda_1$ and $\Lambda_2$ long enough so that the modification induced by a single pulse remains the same, the average photonic potential scales as $V_{\Lambda_1}(x,y) \Lambda_1= V_{\Lambda_2}(x,y) \Lambda_2$. That is, longer periods provide less optical confinement, but conserves the transverse shape of refractive index profile.

Experimentally, we wrote 5~mm long waveguides at a fixed energy $E_\mathrm{{in}} = 0.9~$µJ for different scanning velocities, in turn corresponding to different periods $\Lambda$.
The normalized near-field images are shown in Figure~\ref{fig:wg_periodicity}~(a), together with the respective fitting parameters as a function of $\Lambda$ in Figure~\ref{fig:wg_periodicity}~(b). 
As expected, the confinement drops continuously with the period, whereas the exponential tails get considerably broader. The width of the central Gaussian peak stays nearly constant over the explored range, but the errors increase due to increased noise in the data.
The insertion loss shown in Figure~\ref{fig:wg_periodicity}~(c) undergoes a minimum of 21~dB at $\Lambda=2.07~$µm. For longer periods, the guided mode becomes leaky, whereas for shorter periods the percentage of modified material, and in turn scattering centers, gets higher. Further, at the smallest period a minor overlap of the individual zones might be present, influencing the inscription process.
To model these waveguides, we considered a structure with a square wave longitudinal modulation. The different period is accounted for by varying the duty cycle, but keeping the positive waveform constant. To be consistent with the previous simulations, we considered a duty cycle of $50\%$ when the period is 2~µm. The transverse distribution is W-shaped with $w_\mathrm{guide}=2.85~$µm, with an amplitude such to provide $V_0=6.5\times 10^{-3}$ for the averaged potential. Simulations in Figure~\ref{fig:wg_periodicity}~(d) show the output profile for different periods $\Lambda$. In agreement with the experiments, for $\Lambda>2~$µm the output becomes much wider, yielding an overall increase of the measured insertion loss. Further, the coupling reduces as shown in the inset, thus the propagation loss does not vary significantly.

\subsubsection{\label{sec:shift}Shift experiment}

\begin{figure*}
\includegraphics[width=15cm]{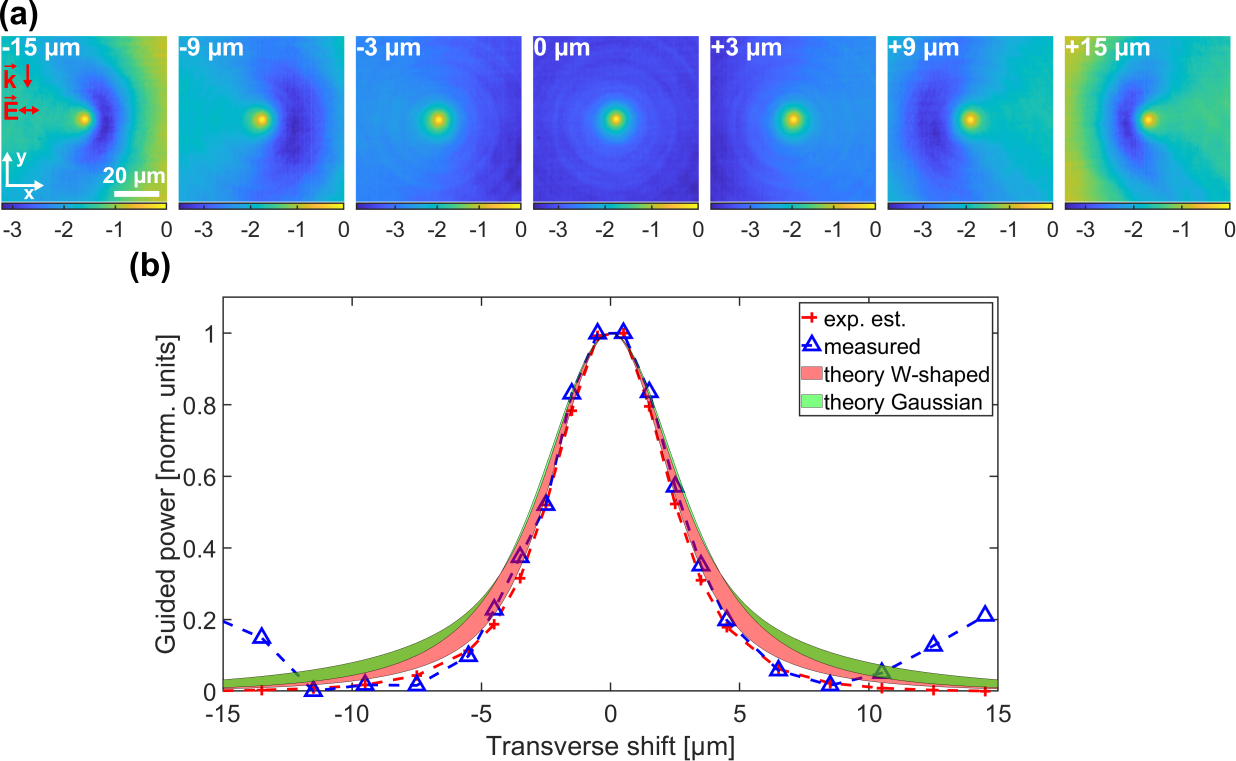}
\caption{\label{fig:shift_result} (a) Normalized logarithmic near field images for different shifted input positions as indicated for a waveguide written with 0.9~µJ and 2~µm pulse-to-pulse pitch. (b) Comparison between the normalized integrated power of the experimental images (blue) and a calculated output power for a shifted input by calculation of the overlap integral (red). The red (green) shaded area corresponds to the theoretical results assuming a W-shape (Gaussian) refractive index profile}
\end{figure*}

Up to this point, the input beam $\varphi(x,y)$ was aligned with sub-micron precision to the waveguide center to optimize the coupling. By slightly shifting the input beam by a transverse shift $x_0$, one can probe the guiding structures in a different manner. Given that now the parity symmetry is broken at the entrance facet, we can examine if guided modes of order $m,l>0$ are also supported by the waveguide. Indeed, if the waveguide is long enough to disperse away the radiation modes, the captured output conserves its profile with $x_0$ only when the waveguide is monomodal. Acquired output profiles for different shifts $x_0$ are shown in Figure~\ref{fig:shift_result}~(a) for a waveguide written with 0.9~µJ pulse energy and a period of 2~µm. To emphasize the radiation modes, normalized logarithmic near-field images are shown, for which the offsets $x_0$ span the range $-15~$µm$~\leq x_0\leq15~$µm. Although changing its amplitude, the shape of the profile around the waveguide core (in the center of the image) does not vary significantly, thus confirming the mono-modality. The amount of power coupled to the fundamental mode provides an additional information on the waveguide. Calling $P_\mathrm{{in}}$ and $P_\mathrm{{guided}}$ the input power and the power coupled to the fundamental mode, respectively, the overlap integral provides
\begin{equation}
\eta(x_0) = \frac{P_\mathrm{{guided}}}{P_\mathrm{{in}}} = \frac{ \left| \int{\varphi(x-x_0,y) \psi_{00}(x,y) dxdy} \right|^2}{{\int |\varphi|^2dxdy} {\int |\psi_{00}|^2dxdy}}.
\end{equation}
The behavior of $\eta(x_0)$ is displayed in Figure~\ref{fig:shift_result}~(b). The guided power $P_\mathrm{{guided}}$ is measured by integrating the camera images (blue dashed line with triangles) in a rectangle of $74 \times 74$~µm$^2$ around the core. The red dashed curve is a prediction where the measured input beam is artificially shifted by $x_0$ and then multiplied by the mode profile measured in the case $x_0=0$. A minor offset of 750 nanometer was incorporated numerically to account for a slightly shifted center position.  
The excellent agreement between the experimental data points and the prediction proves once again that, even though the input is shifted, the same fundamental mode is excited. Additionally, this confirms that the observed mode in this case is not leaky. The deviation between the curves for the largest shifts can be attributed to collected stray light transmitted outside of the waveguide, i.e., the diffraction modes are not negligible. Theoretical computation derived from the overlap integral are reported in the same figure and correspond to the shaded areas. Although the predictions for W-shaped guides (red color) is in better agreement with the experimental results than for the Gaussian ones (green color), the differences between the two sets of curves is too small to provide a definitive evidence. In both cases, the theoretical calculations slightly overestimate the experimental data for large shifts, suggesting the major differences arise from the tails of the field. The easier explanation is that the real waveguide does not match neither Equation~\eqref{eq:gaussian} nor Equation~\eqref{eq:Wshape}, but it is a similar shape which cannot be expressed in a closed form.

\section{Conclusion}
In conclusion, transverse inscription of segmented waveguides in silicon with picosecond laser pulses has been shown. Pulses with a duration of 43~ps enable deterministic subsurface single-pulse modifications. Modifications with separations of around 2~µm lead to repeatable periodic structures showing guiding properties.  A large inscription energy window for which the resulting structures show fundamental mode guidance in a radially symmetric way is found. Weak confinement and a pronounced surrounding region of exponential decay is observed for waveguides written with low pulse energies. The confinement increases with increased pulse energies, the insertion loss exhibit a local minimum for intermediate energies of 20~dB for 4.5~mm long waveguides. The corresponding propagation loss amounts to 2.3 dB/mm. The measured losses depend on the polarization of the probing beam. While inscription with low pulse energy shows similar losses for two orthogonal linear polarization cases, the difference for the highest energy amounts to 6.8~dB. A maximum refractive index change of around $7.5 \times 10^{-3}$ is determined by the inversion of the Helmholtz equation for the near-field images of structures written with the highest energies. The calculated effective refractive index $n_\mathrm{{eff}}$ of the structure increases from pulse energies of >0.7~µJ upwards and reach a maximum value of $1.5 \times 10^{-4}$. Numerical simulations are in excellent agreement with the experiments, providing very similar output fields for waveguides, width, and amplitude compatible with the measurements. The data suggest the presence of a depressed-index region surrounding the core of the waveguide, but more precise measurements are required to confirm the existence of such a region. The confinement can be further controlled by changing the period between the single-pulse modifications.

Overall, this work can be seen as a first step towards the integration of fully three-dimensional optical components in the bulk of silicon with sub-nanosecond laser pulses. 

\section{Methods}
\textit{Inscription setup}: The inscription setup is described in detail in the Supporting Information, Section I. Briefly, it consists of an Er-doped fiber laser system (Raydiance Inc., Smart Light 50) delivering pulses of 860~fs pulse duration at 1.55 µm wavelength with a variable repetition rate. A self-build pulse stretcher is used to increase the pulse duration to 43~ps to avoid catastrophic nonlinearities and reach a deterministic regime for the production of modifications.\cite{Das2020} An aspheric lens (Thorlabs, C037TME-D) with numerical aperture NA = 0.85 is used to focus the beam into the crystalline silicon sample (Sil'tronix Silicon Technologies, \textless 100\textgreater~surface orientation, double side polished, undoped \textgreater 200 $\Omega$ cm$^{-1}$, 500~µm thickness). Shadowgraphy in-situ observation of internal modifications is realized and used to ensure a deterministic modification process for the chosen parameters. A fixed laser repetition rate of 150~Hz combined with a scanning velocity of 0.3~mm/s leads to a pulse-to-pulse pitch of 2~µm.

\textit{Inscription strategy}: At first, a broad range of focusing depth, pulse energy, and pulse-to-pulse pitch was explored to find appropriate conditions for deterministic single-pulse modifications. Circular and linear polarization along and perpendicular to the inscription direction were employed. A preliminary investigation showed waveguiding ability in a wide parameter window in terms of pulse energy for linear polarization along the writing direction. Perpendicular and circular polarization showed guiding possibility only at pulse energies close to the modification threshold. To attain robustness towards unavoidable small variations for the inscription conditions due to pulse-to-pulse energy variance of the laser (<1\% according to manufacturer) or local defects and dust on the surface, only waveguides written with linear polarization along the inscription direction are characterized in detail. Due to the use of an aspheric lens with a design wavelength (9~µm) far away from the employed one (1.55~µm), strong spherical aberrations are to be expected, especially when working in the bulk of silicon which exhibits a high refractive index ($n \approx$ 3.5). To ease the finding of appropriate conditions a constant focal depth of 200~µm below the surface was employed for this study. This depth was chosen because modifications could be induced here within a wide range of energies without damaging the entrance surface of the sample. The pulse-to-pulse pitch was found to be optimal around 2~µm.

\textit{Estimated critical power}: The ratio $P/P_\mathrm{{cr}}$ ranges from 0.31 to 0.79 for the employed pulse energies with Fresnel-reflection taken into consideration, where $P$ is the pulse-peak-power defined by $P = 0.94\times E_\mathrm{{in}} / \tau$ and $P_\mathrm{{cr}}$ is the critical power for self-focusing $P_\mathrm{{cr}} = 3.72 \lambda^2 / ( 8\pi n_0 n_2 ) = $ 24.8 kW. Here, $\lambda$ is the wavelength, $n_0$ is the refractive index, and $n_2$ is the nonlinear refractive index. The value for $n_2$ is taken as $4.1 \times 10^{-18}$ m$^2$ W$^{-1}$.\cite{Dinu2003,Bristow2007,Lin2007} 

\textit{Transmission microscopy}: Infrared shadowgraphy images were taken by a self-build microscopy setup consisting of a super-luminescent laser diode (Thorlabs, S5FC1018P) emitting light at a wavelength of 1317~nm, an NA = 0.85 objective lens (Olympus, LCPLN100XIR), and an InGaAs camera (Xenics, Bobcat320).

\textit{Raman spectroscopy}: A commercial Raman spectrometer (Renishaw, inVia) was used in mapping mode. A 10$\times$10~µm$^2$ area containing the cross section of the waveguide was measured in 250~nm steps in $x$ and $y$ direction. The employed laser of 532~nm wavelength was focused by an objective with numerical aperture NA = 0.85. 

\textit{Waveguide characterization measurement}: The waveguides are characterized in an end-fire configuration described elsewhere and detailed more extensively in the Supporting Information, Section I.\cite{Alberucci2020} Briefly, light at 1.55 µm wavelength is coupled to the waveguides by a 20$\times$ objective lens and is detected by a microscope setup with a 50$\times$ objective lens. An InGaAs camera is used in combination with calibrated neutral-density filters to measure the transmitted power.

\textit{Interpolating function for the guided mode}: The interpolating function $F$ given by the convolution between a Gaussian and a bilateral exponential can be expressed in terms of special functions. A direct computation provides $F(x)\propto e^{-\frac{x}{l}} \text{erfc} \left[\frac{\sqrt{2}}{w} \left(\frac{w^2}{4l}-x\right)  \right] \nonumber  + e^{\frac{x}{l}}\text{erfc} \left[\frac{\sqrt{2}}{w} \left(x+\frac{w^2}{4l}\right) \right] $, where $\text{erfc}$ is the complementary error function.

\textit{Numerical methods}: The optical propagation is computed using a home-made beam propagation method (BPM) based upon operator splitting, with the Crank-Nicolson algorithm used to solve the diffractive term. We also set a vanishing field at the largest distance considered in the numerical scheme. A super-Gaussian attenuator is inserted to dampen out the back reflections. On the other side, we set the first derivative of the field to be vanishing at the origin $r=0$. We employed log-polar coordinates transforming the radial distance as $r=e^{\rho}$. Due to the coordinate transformation, the numerical grid is much denser in proximity of the origin $r=0$, thus drastically increasing the accuracy. The eigenvalue problem is solved analogously, that is, using a finite-difference in the log-polar coordinate for the time-independent Schr\"{o}dinger equation, and then finding numerically the eigenvectors and eigenvalues of such an operator.

\begin{acknowledgments}
We wish to acknowledge the work of C. Otto (Institute of Applied Physics, University Jena) for sample cutting and polishing. M.B. thanks TRUMPF GmbH + Co. KG for financial support. M.B. and M.C. acknowledge funding by the German Federal Ministry of Education and Research through project RUBIN-UKPiño (Grant no. 03RU2U033H). N.A. is part of the Max Planck School of Photonics supported by BMBF, Max Planck Society, and Fraunhofer Society.
Open Access funding enabled and organized by Projekt DEAL.

\end{acknowledgments}

\section*{Data Availability Statement}

The data that support the findings of this study are available from the corresponding author upon reasonable request.

\appendix

\section*{References}

\bibliography{mendeley_export}

\end{document}